\documentclass[11pt]{article}
\usepackage{epsf} 
\usepackage{amsmath}
\usepackage{amssymb}
\usepackage{epsfig}
\usepackage{latexsym}
\usepackage{amsfonts}
\usepackage{graphicx}
\usepackage{cite}
\usepackage{color}

\begin{document}
\begin{center}	
\Large{QUARK-HADRON DUALITY: PINCHED KERNEL APPROACH}
\end{center}
\vspace{.05cm}
\begin{center}
	{\bf  C. A. Dominguez}$^{(a)}$, 
	{\bf L. A. Hernandez}$^{(a)}$, 
	{\bf K. Schilcher}$^{(a),(b)}$,
	{\bf H. Spiesberger} $^{(a),(b)}$  \\ 
\end{center}

\begin{center}
	{\it $^{(a)}$Centre for Theoretical and Mathematical Physics, 
		and Department of Physics, University of
		Cape Town, Rondebosch 7700, South Africa}\\
	
	{\it $^{(b)}$PRISMA Cluster of Excellence, Institut f\"{u}r Physik, 
		Johannes Gutenberg-Universit\"{a}t, D-55099 Mainz, Germany}
	\\
\end{center}

\begin {abstract}
\noindent
Hadronic spectral functions measured by the ALEPH collaboration in the vector and axial-vector channels are used to study potential quark-hadron duality violations (DV). This is done entirely in the framework of pinched kernel finite energy sum rules (FESR), i.e. in a model independent fashion.
The kinematical range of the ALEPH data is effectively extended up to $s = 10\; {\mbox{GeV}^2}$ by using an appropriate kernel, and assuming that in this region the spectral functions are given by perturbative QCD. Support for this assumption is obtained by using $e^+ e^-$ annihilation data in the vector channel. Results in both channels show a good saturation of the pinched FESR, without further need of explicit models of DV.
\end{abstract}

\section{Introduction}
In this talk I address the issue of (global) duality violations (DV) in QCD sum rules \cite{SVZ,QCDSRreview}, particularly in Finite Energy QCDSR (FESR), first proposed in \cite{Shankar}. The formulation is made in the complex squared energy, $s$-plane, where hadronic singularities are present along the positive real axis, and Cauchy's theorem is invoked. Thus, a relation is obtained between hadronic physics on the real axis, and QCD on the circle of radius $s_0$. The latter must be large enough for QCD to be valid. The first problem with DV is that they are difficult to
estimate because they are unknown by definition,
as emphasized in \cite{Pich1}. There are two major approaches to this subject, (i) designing specific models of DV \cite{Pich1,Pich4}, and (ii) 
assuming that global duality is valid on the circle of radius 
$s_{0}$ in the $s$-plane, away from
the positive real axis \cite{PINCH1, PINCH2}. The latter alternative requires only pinched kernels, thus it is basically model independent. Pinching 
serves two purposes, i.e.\ it reduces the contribution of the contour 
integral in the vicinity of the cut, and it can be chosen so as to reduce 
the experimental uncertainties in the integral over the spectral 
function near the kinematic limit of $\tau$-decay.
\\
The approach I will discuss is an extension/improvement over a pinched kernel procedure proposed some time ago \cite{CAD2}. The idea is to choose a suitable pinched integration kernel allowing to perform a FESR 
analysis beyond the kinematical end-point of the $\tau$-decay data, e.g.\ 
up to $s\simeq10\,\mbox{GeV}^{2}$.  The expectation is that in this extended region the non-existing data would be described
by PQCD. In the vector-current channel this expectation can be fully tested using data from $e^+ e^-$ annihilation.\\
The starting point is the definition of the vector and the axial-vector correlators

\begin{align}
\Pi^{VV}_{\mu\nu}(q) &= i\int d^{4}x \,e^{i q x}\,\left\langle 0|T(V_\mu(x)
V_\nu^{\dagger}(0))|0\right\rangle \nonumber \\
&= (-g^{\mu\nu} q^{2} + q^{\mu} q^{\nu}) \Pi_V(q^2)\;, 
\label{Eq:01}
\end{align}

\begin{align}
\Pi^{AA}_{\mu\nu}(q) &= i\int d^{4}x \,e^{i q x}\,\left\langle 0|T(A_\mu(x)
A_\nu^{\dagger}(0))|0\right\rangle \nonumber \\
&= (-g^{\mu\nu} q^{2} + q^{\mu} q^{\nu}) \Pi_A^{(1)}(q^2) + q^{\mu} q^{\nu}\Pi_A^{(0)} (q^2) \;, 
\label{Eq:02}
\end{align}

\noindent
where  $V^{\mu}(x)= \bar{d}(x) \gamma^{\mu}u(x)$, and $A^{\mu}(x)= \bar{d}(x) \gamma^{\mu}\gamma_{5}u(x)$. The PQCD spectral functions in the chiral limit are normalized as  
$\operatorname{Im} \Pi_V|_{PQCD}(s) \equiv \operatorname{Im} \Pi_A|_{PQCD}(s)=\frac{1}{4\pi} \left[ 1 + {\cal{O}}(\alpha_s)\right] \;.$

\noindent
Invoking Cauchy's theorem in the complex square energy $s$-plane leads to the FESR

\begin{equation}
\int_{0}^{s_0}ds\;P(s)\frac{1}{\pi}\operatorname{Im}%
\Pi_{V,A}^{HAD}\left(s\right) = \;\oint_{|s|=s_{0}}ds \,P(s) \,\Pi_{V,A}^{QCD}(s) \;,
\label{Eq:03}
\end{equation}

\noindent
where $P(s)$ is an analytic integration kernel. By requiring that $P(s)$ vanish at some point on the positive real axis one expects to quench potential DV \cite{PINCH1,PINCH2}. An example of a successful application of this method is the Weinberg sum rules \cite{PINCH2}. In fact, the simple pinched kernel 

\begin{equation}
P_0(s)= 1 - \frac{s}{s_0} \;,
\label{Eq:04}
\end{equation}

\noindent
results in the saturation of these sum rules, something which does not happen if $P_0(s)=1$.

\section{Pinched QCD Sum Rules}
In reference\cite{CAD2} the FESR range was effectively extended from the kinematical end-point of the $\tau$-decay data, $s = s_1 \, \approx \, M_\tau^2$ to $s = s_0 \, \approx 10\; {\mbox{GeV}^2}$ by assuming that the spectral function was a constant for $s \ge s_1$, and $P(s)$ subject to the quenching constraint

\begin{equation}
	\int_{s_{1}}^{s_{0}}P_0(s)\,ds=0.
\label{Eq:05}
\end{equation}

\noindent
This assumption can be relaxed and simply require that for $s > s_1$ the spectral function be given by PQCD, a more likely scenario. After searching for the optimal $P(s)$ we found the linear function to be the most suitable, i.e.

\begin{equation}
P_1(s)=1+\gamma(s_0, s_1)\, s \;,
\label{Eq:06}
\end{equation}

\noindent
where the function $\gamma(s_0, s_1)$ is found to be

\begin{equation}
\gamma(s_0,s_1) = - \,  \frac{s_0 M_0(s_0) - s_1 M_0(s_1)}{s_0^2 M_1(s_0) - s_1^2 M_1(s_1)}\,,
\end{equation}

\noindent
where the dimensionless moments, $M_N(s_0)$, are

\begin{eqnarray}
M_{N}(s_0) &=& \frac{4\,\pi^2}{s_0^{(N+1)}} \, 
\int_0^{s_0} ds \; s^N \; \frac{1}{\pi} \, {\mbox{Im}} \, \Pi^{PQCD}(s)
\nonumber \\[.3cm]
&=& - \frac{1}{2 \pi i} \; \frac{4\pi^2}{s_0^{N+1}} \;\oint_{|s|=s_0} \; ds \;\Pi^{PQCD}(s)
\; , 
\label{Eq:7bis}
\end{eqnarray}

\noindent 
The FESR in the axial-vector and the vector channel become

\begin{equation}
2 \,f_{\pi}^{2} = - \int_{0}^{s_{1}} ds \,P_1(s)\,\frac{1}{\pi
} \,\operatorname{Im}\Pi_{A}^{DATA}\left(s\right)  - \frac{1}{2\pi i}
\,\oint_{|s|=s_{0}}ds \,P_1(s) \,\Pi_{A}^{PQCD}(s) \;,\label{Eq:08}
\end{equation}

\begin{equation}
F(s_0) = 
- \int_0^{s_1} ds \, P_1(s)\, \frac{1}{\pi}\, {\mbox{Im}} \Pi_V^{DATA}(s) 
- \frac{1}{2\pi i} \, \oint_{|s|=s_{0}} ds \, P_1(s) \, \Pi(s)_V^{PQCD}(s)\;, 
\label{Eq:09}
\end{equation}

\noindent
where the contribution from the QCD vacuum condensates in the OPE  is negligible (only the dimension-four gluon condensate contributes in this case). The left-hand-side of Eq.(9), $F(s_0)$, is actually zero (no pole in this channel).
\begin{figure}
	[ht]
	\begin{center}
		\includegraphics[width=4.0in]{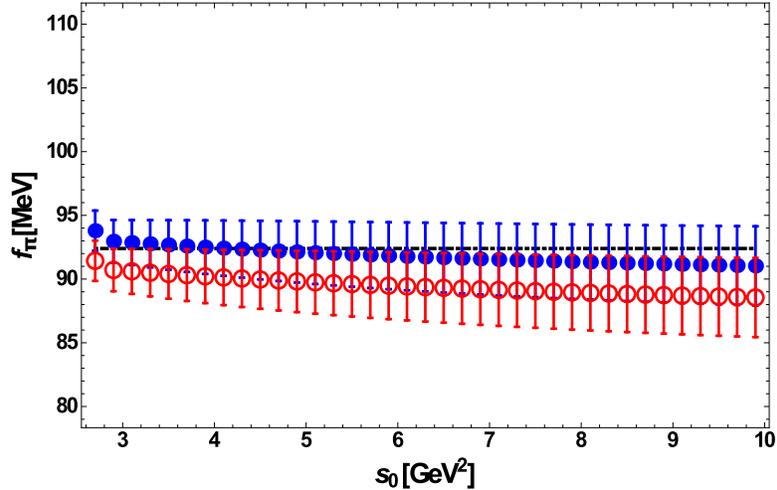}
		\caption{\footnotesize{Results for $f_\pi$ as a function of 
		$s_0$ from the FESR, Eq.\ (\ref{Eq:08}), for the pinched kernel $P_1(s)$,
		Eq.\ (\ref{Eq:06}), with $s_1 = 2.7 \, {\mbox{GeV}^2}$. 
		The corrected ALEPH data \cite{ALEPHnew} has been used in the 
		hadronic integral. Solid circles are for $\alpha_s = 0.354$, 
		and open circles for $\alpha_s= 0.328$, corresponding to the 
		maximum and minimum values of $\alpha_s = 0.341 \pm 0.013$ from 		\cite{Pich3}. }}
	\end{center}
\end{figure}

\begin{figure}
	[ht!]
	\begin{center}
		\includegraphics[width=4.3in]{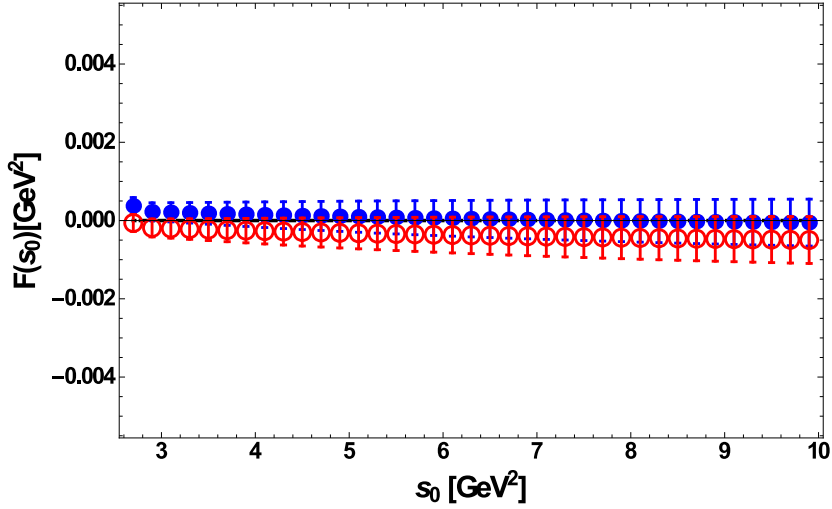}
		\caption{\footnotesize{ 
		Results in the vector channel, i.e.\ $F(s_0)$ as a function of 
		$s_0$ from the FESR, Eq.\ (\ref{Eq:09}), for the pinched kernel $P_1(s)$,
		Eq.\ (\ref{Eq:06}), with $s_1 = 2.7 \, {\mbox{GeV}^2}$. The  
		corrected ALEPH data \cite{ALEPHnew} has been used in the 
		hadronic integral. The values of $\alpha_s$ are the same as in Fig. 1. }}
	\end{center}
\end{figure}

\noindent
The result for $f_\pi$ as a function of $s_0$, with $\alpha_{s}(m_{\tau}^{2})=0.341 \pm 
0.013$  \cite{Pich3} is shown in Fig.\ 1. There is good agreement 
between the right-hand-side of this FESR and the experimental value 
of $f_\pi$. Similar agreement is obtained in the vector channel, as shown in Fig.2.\\

\begin{figure}
	[t!]
	\begin{center}
		\includegraphics[width=4.0in]{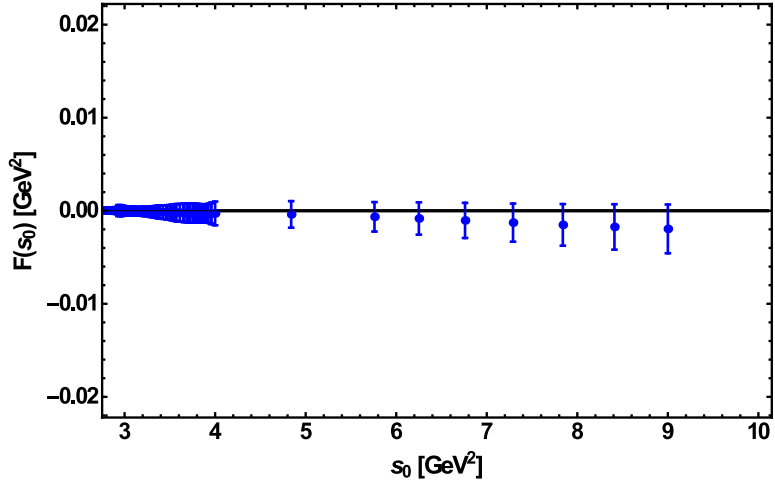}
		\caption{\footnotesize {Results in the vector channel, i.e.\  
		$F(s_0)$ as a function of $s_0$ from the FESR Eq.\ (\ref{Eq:10}), 
		with the pinched kernel $P_0(s)$, Eq.\ (\ref{Eq:04}). The corrected ALEPH 
		data \cite{ALEPHnew} has been used followed by $e^+ e^-$ data \cite{eeData} beyond the $\tau$-decay kinematical
		end point.}}
	\end{center}
\end{figure}

\noindent
The assumption made in the vector channel regarding the region above the $\tau$-decay kinematical end point can be tested by using actual data from $e^+ e^-$ annihilation into hadrons. The sum rule Eq. (9) becomes

\begin{eqnarray}
F(s_0) 
&=& 
- \int_0^{s_1} ds\,P(s)  \, 
\frac{1}{\pi}\, {\mbox{Im}} \,\Pi_V^{\tau}(s) 
- \frac{1}{8\, \pi^2} \int_{s_1}^{s_0} ds\,P(s) \, R(s) 
\nonumber \\[.3cm]
& & 
- \frac{1}{2 \pi i} \oint_{C(|s_0|)} ds \, P(s) \,\Pi_V^{PQCD}(s) 
\,,
\label{Eq:10}
\end{eqnarray}

\noindent
where the first integral involves the ALEPH data, and the second integral requires the $e^+ e^-$ data on the ratio $R(s)$, i.e. $R(s) = 8 \, \pi \, {\mbox{Im}} \, \Pi_V(s)$.
It is important to compare Fig.2, obtained by assuming PQCD in the extended region, with Fig.3, for which actual $e^+ e^-$ data was used instead.
In addition, the pinched kernel $P_1(s)$, Eq.(6), was used in Fig. 2, and $P_0(s)$, Eq.(4), in Fig.3. The reason being that since $e^+ e^-$ data is being used in the extended region, there is no need to require the spectral function to be given by PQCD.
The central values in Fig.3, while agreeing with the expected
result $F(s_0)=0$, suggest  a small systematic deviation from the horizontal. This deviation increases with increasing energy. This behaviour may be traced back to the behaviour  of $R(s)$ in this energy region, as shown on page 535 of \cite{PDG}. Indeed, the $e^+ e^-$ data lie systematically above PQCD in the region $2.0 \, {\mbox{GeV}} \lesssim\sqrt{s} \lesssim 
3.0 \, {\mbox{GeV}}$. If one were to renormalize these data then the  downward trend would disappear. 
\section{Conclusions}
In this talk I discussed a method to explore an energy region well above the kinematical end point of the ALEPH data on $\tau$-lepton decay, in the vector and the axial-vector channels. The purpose is to ascertain if simple pinched kernels in FESR are sufficient to tame potential quark-hadron global duality violations. The results fully support this alternative to specific DV models which involve a large number of free parameters. Nevertheless, it must be kept in mind that the size of expected DV is channel/application dependent. In this connection it is important to mention a recent determination of chiral couplings and vacuum condensates from FESR using a model of DV \cite{Pich4}. This model involves $3 \times 10^6$-tuples of a set of four parameters. The results are in good agreement with our pinched-kernel determinations in \cite{CAD3,CAD4}.

\section*{Acknowledgements}

We acknowledge financial support by the Deutsche Forschungsgemeinschaft,  the Mainz Institute for Theoretical Physics (MITP), and the National Research Foundation (South Africa).

\end{document}